\documentclass[
reprint,
superscriptaddress,
nobibnotes,
amsmath,amssymb,
aps,
prb,
]{revtex4-2}

\usepackage{bm}
\usepackage{amsmath}
\usepackage{amssymb}
\usepackage{amsthm}
\usepackage{graphicx}
\usepackage{hyperref}

\newtheorem{criterion}{Criterion}
\newtheorem{proposition}{Proposition}

\begin{document}

\title{Thermodynamic incompleteness of state dynamics in Markovian transport}
\thanks{Correspondence should be addressed to Yang Tian.}

\author{Yang Tian}
\email{tyanyang04@gmail.com \& yang.tian@infplane.com}
\affiliation{Infplane Computing Technologies Ltd, Beijing, 100080, China}

\begin{abstract}
Markovian transport is often described by a master equation for the system
state. The thermodynamic information measured in transport experiments,
however, is carried by reservoir-resolved transfer records, such as particle
currents, heat currents, entropy production, and current noise. We identify a
thermodynamic incompleteness of state dynamics: a Markovian state generator can
fix the occupation probabilities, stationary response, and relaxation without
specifying how the underlying transitions are assigned to reservoirs and energy
filters. We study a multi-terminal Coulomb-blockaded quantum dot coupled to
energy-filtered reservoirs, for which different assignments of reservoir
channels can generate the same state master equation. These assignments give
identical occupation dynamics, stationary state, and linear response of the
dot, but different heat currents, entropy production, and current noise. We
formulate a thermodynamic completeness criterion: a transport observable can be
reconstructed from state dynamics only when it is invariant under all changes
of reservoir-channel assignments that leave the state generator unchanged. The
criterion gives a practical diagnostic for Markovian transport models and a
measurable prediction: state tomography can be insufficient to predict
heat-noise and cross-correlation measurements, even when the full Markovian
state dynamics is known. The analysis identifies a concrete limitation of
state-only Markovian thermodynamics and shows which additional transport
records must be specified to make thermodynamic predictions experimentally
complete.
\end{abstract}

\maketitle

\section{Introduction}
\label{Introduction-section}

Markovian master equations are a standard language for transport through
mesoscopic conductors, quantum dots, thermal machines, and monitored open
systems
\cite{GoriniKossakowskiSudarshan1976,Lindblad1976,Davies1974,Spohn1978,EspositoHarbolaMukamel2009,LandiPaternostro2021,StrasbergWinter2021,LandiCurrent2024}. A master equation for the system state
determines relaxation rates, stationary occupations, and state response
functions. Transport experiments often measure reservoir-resolved records
instead, including particle, energy, heat, spin, photon, or other transfer
statistics
\cite{PtaszynskiEsposito2023,LiuCounting2023,HasegawaThermodynamic2021,HasegawaContinuous2020,HasegawaThermodynamic2022,SalazarThermodynamic2024,MonnaiThermodynamic2022,LiuCoherences2021,GerryFull2023,PaulinoDeviation2024,TirritoFull2023,ZhangFull2021,WangDistinguishing2024,FanFull2024,Perfettojump2022}. The
central question of this paper is whether complete knowledge of the Markovian
state dynamics is enough to determine those thermodynamic records.

The issue arises because a state generator contains total transition rates
between system states, whereas a thermodynamic transport model also specifies
which reservoir or energy-filter channel realizes each transition and which
increment is recorded in that channel. We call a state-only Markovian
description thermodynamically incomplete for a given record when it fixes the
state probabilities and their response, but does not fix that record because
different reservoir-channel assignments remain compatible with the same
generator. This mechanism is distinct from hidden-state inference or
coarse-grained non-Markovian dynamics, where missing entropy production is
caused by eliminating part of the state space
\cite{Seifert2019Inference,DeguentherVanDerMeerSeifert2024,MaierDeguentherVanDerMeerSeifert2024,ErtelSeifert2024,HarunariFioreBarato2024,OtsuboManikandanSagawaKrishnamurthy2022,ManikandanGuptaKrishnamurthy2020,Ehrich2021,KapustinGhosalBisker2024}.
Here the state equation itself is assumed to be known exactly. No transition
rate in the state generator is missing. The missing information is the
assignment of those transitions to reservoirs, filters, and measurement
records.

The practical issue is that state tomography is often used to validate a
Markovian transport model before the model is used to predict thermodynamic
performance
\cite{PtaszynskiEsposito2023,LiuCounting2023,Seifert2019Inference,ErtelSeifert2024}.
We show when that inference is justified and when it is
underdetermined: all state-level tests can agree with the same generator,
while heat currents, entropy production, and noise measurements distinguish
the underlying reservoir-channel assignments.

We first formulate a reconstruction criterion: a record observable is
determined by the state generator only if it is unchanged under every
redistribution of transition rates among reservoir channels that keeps the
generator fixed. We then apply the criterion to a multi-terminal
Coulomb-blockaded quantum dot with energy-filtered contacts
\cite{SanchezSothmannJordanButtiker2013,WoznyLeijnse2025,KleinherbersSchuenemannKoenig2023},
where identical
occupation dynamics can coexist with different heat currents, entropy
production, and current noise. Finally, we derive the geometric and
topological form of the criterion. The projection from channel currents to
state dynamics gives a quotient fluctuation geometry, while the connectivity
of the transport channels gives exact bounds on the heat, particle, and spin
records still compatible with the same state generator.

Section \ref{Model-section} defines the transport data and thermodynamic
records. Section \ref{Criterion-section} states the reconstruction criterion.
Section \ref{Dot-section} constructs state-identical quantum-dot devices with
different thermodynamic records. Section \ref{Geometry-section} derives the
geometric and topological form of the criterion. Section
\ref{Conclusion-section} summarizes the experimental implication.

\section{Markovian transport data}
\label{Model-section}

We consider a finite set of system states \(n=1,\ldots,N\). The probability
vector \(p(t)\) obeys the Markovian master equation
\begin{align}
    \frac{\mathsf{d}}{\mathsf{d}t}p_{n}
    =
    \sum_{m\neq n}
    \left[
    W_{nm}p_{m}
    -
    W_{mn}p_{n}
    \right]
    \equiv
    \left(Lp\right)_{n}.
    \label{Eq-state-master}
\end{align}
The generator \(L\) contains the total transition rate
\(W_{nm}\) from state \(m\) to state \(n\). A transport experiment generally
resolves this total rate into reservoir or measurement channels,
\begin{align}
    W_{nm}
    =
    \sum_{\alpha\in{\cal A}_{nm}}
    w_{nm}^{\alpha}.
    \label{Eq-channel-resolution}
\end{align}
The label \(\alpha\) specifies the physical channel that realizes the
transition. In an electronic conductor it can include the reservoir index, an
energy-filter index, or a spin-selective contact. Each channel also carries a
record increment \(d_{nm}^{\alpha,\mu}\) for the measured quantity \(\mu\),
such as transferred charge or reservoir heat.

For a probability vector \(p\), the mean rate of the record \(\mu\) is
\begin{align}
    J_{\mu}(p)
    =
    \sum_{m,n,\alpha}
    d_{nm}^{\alpha,\mu}w_{nm}^{\alpha}p_{m}.
    \label{Eq-mean-current}
\end{align}
The full counting statistics are generated by the tilted Markov generator
\begin{align}
    \left[L(\bm{\chi})\right]_{nm}
    =
    \sum_{\alpha\in{\cal A}_{nm}}
    w_{nm}^{\alpha}
    \exp\left(
    \sum_{\mu}\chi_{\mu}d_{nm}^{\alpha,\mu}
    \right),
    \qquad n\neq m,
    \label{Eq-tilted-offdiag}
\end{align}
with diagonal entries chosen so that \(L(0)=L\). The long-time cumulants are
obtained from the dominant eigenvalue of \(L(\bm{\chi})\)
\cite{EspositoHarbolaMukamel2009,PtaszynskiEsposito2023,LiuCounting2023,WoznyLeijnse2025,SanchezSothmannJordanButtiker2013}.
Eqs. (\ref{Eq-state-master}-\ref{Eq-tilted-offdiag}) show the basic
separation. The state generator depends on \(W_{nm}\). The thermodynamic
records depend on the channel rates \(w_{nm}^{\alpha}\) and on the increments
\(d_{nm}^{\alpha,\mu}\).

\section{Thermodynamic completeness criterion}
\label{Criterion-section}

We now ask when a thermodynamic record is determined by the state generator.
Fix the total rates \(W_{nm}\). A generator-preserving perturbation of the
channel rates is a collection \(\delta w_{nm}^{\alpha}\) satisfying
\begin{align}
    \sum_{\alpha\in{\cal A}_{nm}}
    \delta w_{nm}^{\alpha}
    =
    0
    \quad
    \text{for every transition }m\rightarrow n .
    \label{Eq-generator-preserving}
\end{align}
For sufficiently small amplitude, such a perturbation changes the physical
channel assignment but leaves the Markovian state equation
(\ref{Eq-state-master}) unchanged.

\begin{criterion}
\label{Criterion-completeness}
Consider the Markovian transport model defined by
Eqs. (\ref{Eq-state-master}-\ref{Eq-channel-resolution}). A linear
transport record \(J_{\mu}(p)\) is determined by the state generator \(L\), for
all channel assignments compatible with \(L\), if and only if
\begin{align}
    \sum_{m,n,\alpha}
    d_{nm}^{\alpha,\mu}\delta w_{nm}^{\alpha}p_{m}
    =
    0
    \label{Eq-completeness}
\end{align}
for every probability vector \(p\) and every generator-preserving perturbation
\(\delta w\). If this condition fails, there exist two Markovian transport
models with the same state generator and different values of \(J_{\mu}\).
\end{criterion}

For a generator-preserving channel redistribution, if
Eq. (\ref{Eq-completeness}) holds, changing the channel assignment at fixed
\(L\) does not change the record rate. The record is then a function of the
state generator alone. If Eq. (\ref{Eq-completeness}) fails, one can choose a
small parameter \(\epsilon\) such that
\(w_{nm}^{\alpha}+\epsilon\delta w_{nm}^{\alpha}\geq0\). The perturbed and
unperturbed devices have the same total rates \(W_{nm}\), but their record
rates differ by the nonzero first-order change in
Eq. (\ref{Eq-completeness}). Thus the state generator \(L\) does not contain
enough information to choose a unique reservoir-channel assignment.

The same criterion applies to current noise and higher cumulants. The tilted
generator \(L(\bm{\chi})\) is determined by \(L\) only if every
generator-preserving perturbation leaves the off-diagonal tilted rates in
Eq. (\ref{Eq-tilted-offdiag}) unchanged for the chosen counting fields. If two
channel assignments have the same \(L\) but different \(L(\bm{\chi})\), then
state tomography cannot determine the corresponding counting statistics.

The first two cumulants make the separation between state dynamics and record
data explicit. Let \(\langle{\bf 1}|\) denote the row vector with all entries
equal to one, let \(|p^{\mathrm{ss}}\rangle\) be the stationary state of
\(L\), and define
\begin{align}
    L_{\mu}
    =
    \left.
    \frac{\partial L(\bm{\chi})}{\partial\chi_{\mu}}
    \right|_{\bm{\chi}=0},
    \qquad
    L_{\mu\nu}
    =
    \left.
    \frac{\partial^{2} L(\bm{\chi})}
    {\partial\chi_{\mu}\partial\chi_{\nu}}
    \right|_{\bm{\chi}=0}.
    \label{Eq-tilt-derivatives}
\end{align}
The mean current is
\begin{align}
    J_{\mu}^{\mathrm{ss}}
    =
    \langle{\bf 1}|
    L_{\mu}
    |p^{\mathrm{ss}}\rangle .
    \label{Eq-fcs-mean}
\end{align}
The zero-frequency current-noise matrix can be written as
\begin{align}
    S_{\mu\nu}
    =
    \langle{\bf 1}|L_{\mu\nu}|p^{\mathrm{ss}}\rangle
    -
    \langle{\bf 1}|
    \left(
    L_{\mu}{\cal R}L_{\nu}
    +
    L_{\nu}{\cal R}L_{\mu}
    \right)
    |p^{\mathrm{ss}}\rangle ,
    \label{Eq-fcs-noise}
\end{align}
where \({\cal R}\) is the Drazin inverse of \(L\) on the subspace orthogonal
to the stationary state
\cite{EspositoHarbolaMukamel2009,PtaszynskiEsposito2023,LiuCounting2023}.
The state generator fixes
\(|p^{\mathrm{ss}}\rangle\) and \({\cal R}\), but it does not fix
\(L_{\mu}\) or \(L_{\mu\nu}\). Thus a state-only model may correctly predict
relaxation and stationary occupations while failing to predict current means,
current noise, or cross correlations.

\section{Multi-terminal energy-filtered quantum dot}
\label{Dot-section}

We apply the criterion to a Coulomb-blockaded quantum dot with \(N\) single
particle levels \(\varepsilon_{i}\)
\cite{SanchezSothmannJordanButtiker2013,WoznyLeijnse2025,KleinherbersSchuenemannKoenig2023}. The dot is either empty, denoted by
\(0\), or occupied by one electron in level \(i\). Reservoir \(r\) has chemical
potential \(\mu_{r}\) and temperature \(T_{r}\). Energy-filtered contacts
produce channel rates
\begin{align}
    w_{i0}^{r\lambda}
    =
    \gamma_{i}^{r\lambda}
    f_{r}\left(\varepsilon_{i}\right),
    \qquad
    w_{0i}^{r\lambda}
    =
    \gamma_{i}^{r\lambda}
    \left[
    1-f_{r}\left(\varepsilon_{i}\right)
    \right],
    \label{Eq-dot-rates}
\end{align}
where \(f_{r}\) is the Fermi function and \(\lambda\) labels an energy-filter
channel. The state dynamics depends only on
\begin{align}
    \Gamma_{i}^{+}
    =
    \sum_{r,\lambda}w_{i0}^{r\lambda},
    \qquad
    \Gamma_{i}^{-}
    =
    \sum_{r,\lambda}w_{0i}^{r\lambda}.
    \label{Eq-total-dot-rates}
\end{align}
The occupation equation is
\begin{align}
    \frac{\mathsf{d}}{\mathsf{d}t}p_{i}
    =
    \Gamma_{i}^{+}p_{0}
    -
    \Gamma_{i}^{-}p_{i},
    \qquad
    p_{0}=1-\sum_{i=1}^{N}p_{i}.
    \label{Eq-dot-master}
\end{align}
Thus the entire occupation dynamics, including stationary occupations and
linear response of the dot probabilities, is fixed by
\(\Gamma_{i}^{+}\) and \(\Gamma_{i}^{-}\).
The stationary solution is
\begin{align}
    p_{0}^{\mathrm{ss}}
    =
    \left(
    1+\sum_{i=1}^{N}
    \frac{\Gamma_{i}^{+}}{\Gamma_{i}^{-}}
    \right)^{-1},
    \qquad
    p_{i}^{\mathrm{ss}}
    =
    \frac{\Gamma_{i}^{+}}{\Gamma_{i}^{-}}
    p_{0}^{\mathrm{ss}} .
    \label{Eq-dot-ss}
\end{align}
The linearized relaxation matrix around this stationary state is
\begin{align}
    A_{ij}
    =
    -\Gamma_{i}^{-}\delta_{ij}
    -
    \Gamma_{i}^{+}.
    \label{Eq-dot-response}
\end{align}
The notation means that the second term is independent of the column index
\(j\). Hence every state-only relaxation measurement is determined by the
same set of total rates \(\{\Gamma_{i}^{+},\Gamma_{i}^{-}\}\).

The reservoir heat record is not fixed by these totals. For an electron
entering the dot from reservoir \(r\), the reservoir heat increment is
\(-(\varepsilon_{i}-\mu_{r})\). For an electron leaving the dot into reservoir
\(r\), it is \(+(\varepsilon_{i}-\mu_{r})\). The heat current from reservoir
\(r\) is therefore
\begin{align}
    \dot Q_{r}
    =
    \sum_{i,\lambda}
    \left(\varepsilon_{i}-\mu_{r}\right)
    \left[
    w_{0i}^{r\lambda}p_{i}
    -
    w_{i0}^{r\lambda}p_{0}
    \right].
    \label{Eq-dot-heat}
\end{align}
The total entropy-production rate of the resolved jump process is
\begin{align}
    \dot S_{\mathrm{env}}
    =
    \sum_{i,r,\lambda}
    \left[
    w_{i0}^{r\lambda}p_{0}
    -
    w_{0i}^{r\lambda}p_{i}
    \right]
    \ln
    \frac{
    w_{i0}^{r\lambda}p_{0}
    }{
    w_{0i}^{r\lambda}p_{i}
    } .
    \label{Eq-dot-entropy}
\end{align}
This expression depends on the reservoir-channel rates before they are summed
into \(\Gamma_{i}^{+}\) and \(\Gamma_{i}^{-}\). It is therefore not a function
of the occupation generator alone. At stationarity, where the system entropy
does not change on average, this rate equals the entropy flow into the
reservoirs.

State data can nevertheless give bounds when the set of possible reservoirs is
known. Let \(q_{ir}=\varepsilon_{i}-\mu_{r}\). Suppose first that an
electron-entering transition \(0\rightarrow i\) can be supplied by reservoirs
in a set \(R_{i}^{+}\), while the total entering rate \(\Gamma_{i}^{+}\) is
known from state tomography. Since the channel rates form a nonnegative
decomposition of \(\Gamma_{i}^{+}\), the entering contribution to the heat
current lies in the interval
\begin{align}
    -p_{0}\Gamma_{i}^{+}
    \max_{r\in R_{i}^{+}}q_{ir}
    \leq
    -p_{0}
    \sum_{r,\lambda}q_{ir}w_{i0}^{r\lambda}
    \leq
    -p_{0}\Gamma_{i}^{+}
    \min_{r\in R_{i}^{+}}q_{ir}.
    \label{Eq-heat-bound-in}
\end{align}
Similarly, if the electron-leaving transition \(i\rightarrow0\) can empty into
reservoirs in \(R_{i}^{-}\), then
\begin{align}
    p_{i}\Gamma_{i}^{-}
    \min_{r\in R_{i}^{-}}q_{ir}
    \leq
    p_{i}
    \sum_{r,\lambda}q_{ir}w_{0i}^{r\lambda}
    \leq
    p_{i}\Gamma_{i}^{-}
    \max_{r\in R_{i}^{-}}q_{ir}.
    \label{Eq-heat-bound-out}
\end{align}
The intervals collapse to single values only when all allowed reservoirs have
the same heat increment for that transition, or when the channel assignment is
specified independently. Thus state dynamics can bound heat currents, but it
does not determine them except in thermodynamically complete cases.

We now give an explicit pair of state-identical devices. They have the same
occupation master equation, but they differ in heat currents and heat-current
noise because the same total dot transition rate is assigned to different
reservoirs.

\begin{proposition}
\label{Proposition-dot}
Assume that at least one dot transition is coupled to two reservoirs \(r\) and
\(s\) with different heat increments,
\(\varepsilon_{i}-\mu_{r}\neq\varepsilon_{i}-\mu_{s}\). Then there exist two
energy-filtered quantum-dot devices with identical occupation master equation
(\ref{Eq-dot-master}) and different reservoir heat currents. Their tilted
generators for heat counting also differ. Hence heat-current noise and heat
cross correlations are not determined by the occupation dynamics alone.
\end{proposition}

To prove the statement, choose a transition \(0\rightarrow i\) and shift a
small rate \(\eta\) from reservoir \(s\) to reservoir \(r\):
\(\delta w_{i0}^{r}=\eta\), \(\delta w_{i0}^{s}=-\eta\), with all other
channels fixed. Eq. (\ref{Eq-total-dot-rates}) is unchanged, so the
occupation dynamics is identical. Eq. (\ref{Eq-dot-heat}) changes by
\(-\eta(\mu_{s}-\mu_{r})p_{0}\). The off-diagonal tilted rate for this
transition changes from
\(\sum_{a=r,s}w_{i0}^{a}e^{-\chi(\varepsilon_{i}-\mu_{a})}\) to the same
expression plus
\(\eta[e^{-\chi(\varepsilon_{i}-\mu_{r})}-e^{-\chi(\varepsilon_{i}-\mu_{s})}]\).
Thus the heat counting statistics differ for generic \(\chi\).

The same conclusion can be expressed directly in the full-counting-statistics
generator. If \(\chi_{r}^{Q}\) counts heat entering reservoir \(r\), the
off-diagonal entries of the tilted dot generator are
\begin{align}
    \left[L(\bm{\chi}^{Q})\right]_{i0}
    =
    \sum_{r,\lambda}
    w_{i0}^{r\lambda}
    e^{-\chi_{r}^{Q}\left(\varepsilon_{i}-\mu_{r}\right)},
    \label{Eq-dot-tilt-in}
\end{align}
\begin{align}
    \left[L(\bm{\chi}^{Q})\right]_{0i}
    =
    \sum_{r,\lambda}
    w_{0i}^{r\lambda}
    e^{+\chi_{r}^{Q}\left(\varepsilon_{i}-\mu_{r}\right)} .
    \label{Eq-dot-tilt-out}
\end{align}
The state generator is recovered at \(\bm{\chi}^{Q}=0\), where only the totals
in Eq. (\ref{Eq-total-dot-rates}) remain. Derivatives with respect to
\(\bm{\chi}^{Q}\) restore the reservoir-resolved heat increments. The dominant
eigenvalue \(\lambda_{\max}(\bm{\chi}^{Q})\) therefore contains information
that is absent from \(L\). In particular, the heat-noise matrix
\begin{align}
    S_{rs}^{Q}
    =
    \left.
    \frac{\partial^{2}\lambda_{\max}}
    {\partial\chi_{r}^{Q}\partial\chi_{s}^{Q}}
    \right|_{\bm{\chi}^{Q}=0}
    \label{Eq-dot-heat-noise}
\end{align}
is not fixed by occupation dynamics when different reservoir assignments give
different tilted generators.

The prediction is experimentally concrete. Two samples can be tuned to have
the same occupation relaxation rates and the same stationary dot populations.
If the tunnel couplings distribute the same total rates among reservoirs with
different chemical potentials or different energy filters, heat-noise and
cross-correlation measurements distinguish the samples even though state
tomography does not.

The channel-space mechanism can be seen explicitly. Consider the two
entering channels \(0\rightarrow i\) from reservoirs \(r\) and \(s\). In the
channel-current basis \((j_{i0}^{r},j_{i0}^{s})\), the projection \(P\) from
reservoir-resolved channels to the total ordered transition current has the
local block
\begin{align}
    P_{i0}^{(r,s)}
    =
    \begin{pmatrix}
    1 & 1
    \end{pmatrix},
    \label{Eq-dot-P-block}
\end{align}
where \(P_{i0}^{(r,s)}j=j_{i0}^{r}+j_{i0}^{s}\). The vector
\begin{align}
    c_{i0}^{rs}
    =
    \begin{pmatrix}
    1\\
    -1
    \end{pmatrix}
    \label{Eq-dot-kernel-vector}
\end{align}
therefore satisfies \(P_{i0}^{(r,s)}c_{i0}^{rs}=0\). It redistributes current
between two reservoir channels without changing the total transition rate, and
therefore without changing the occupation generator. The
heat-record row for these two channels is
\begin{align}
    D_{Q}^{(r,s)}
    =
    \begin{pmatrix}
    -\left(\varepsilon_{i}-\mu_{r}\right)
    &
    -\left(\varepsilon_{i}-\mu_{s}\right)
    \end{pmatrix},
    \label{Eq-dot-D-row}
\end{align}
and hence
\begin{align}
    D_{Q}^{(r,s)}c_{i0}^{rs}
    =
    \mu_{r}-\mu_{s}.
    \label{Eq-dot-Dc}
\end{align}
Whenever the two reservoirs have different chemical potentials, this
generator-invisible channel direction carries a heat record. This is the
finite-dimensional geometric origin of the incompleteness in the dot.

The construction also shows that the effect appears directly in transport
statistics. The matrices \(L_{\mu}\) and
\(L_{\mu\nu}\) entering Eqs. (\ref{Eq-fcs-mean}-\ref{Eq-fcs-noise}) are
changed at first order by a redistribution of rates between reservoirs with
different heat increments. The matrix \(L\), its stationary state, and its
relaxation spectrum are unchanged. Therefore the discrepancy appears directly
in the transport statistics rather than through an error in the inferred state
dynamics. In this regime, state measurements are internally consistent but
thermodynamically incomplete.

\section{Geometric and topological form of the criterion}
\label{Geometry-section}

The criterion above can be expressed in geometric form. We use a directed
transition-channel representation. The nodes are system states, and the
directed channels are reservoir- or filter-resolved transitions. Several
distinct channels may connect the same ordered pair of states. There are two
linear projections. The first projection \(P\) maps channel currents \(j\) to
the total ordered-transition currents \(u=Pj\), with
\(u_{nm}=\sum_{\alpha}j_{nm}^{\alpha}\). The Markovian state generator fixes
this first projection. The second projection \(B\) maps ordered-transition
currents to state velocities, so that \(\dot p=Bu=BPj\). We assume that the
full state generator is known. The primary hidden space is therefore
\(\ker P\), rather than the larger nullspace of \(BP\).

Near a typical current \(\bar j\), the channel-current action has the
quadratic expansion
\begin{align}
    I(j)
    =
    \frac{1}{2}
    \left(j-\bar j\right)^{T}
    R
    \left(j-\bar j\right)
    +o\left(\|j-\bar j\|^{2}\right),
    \label{Eq-current-geometry}
\end{align}
where \(R\) is the local inverse covariance of channel currents. If only the
generator-visible ordered-transition current \(u=P(j-\bar j)\) is retained,
the effective cost is the minimum channel-current cost over all
reservoir-channel perturbations producing the same \(u\):
\begin{align}
    I_{\mathrm{gen}}(u)
    =
    \inf_{P\delta j=u}
    \frac{1}{2}
    \delta j^{T}R\delta j
    =
    \frac{1}{2}u^{T}
    \left(PR^{-1}P^{T}\right)^{\dagger}u .
    \label{Eq-quotient}
\end{align}
The dagger denotes the Moore-Penrose inverse on the range of \(P\). This
quotient geometry is the part of channel-current fluctuations visible in the
state generator. It removes the nullspace \(\ker P\), which consists of
reservoir-channel redistributions that leave all total transition rates
unchanged
\cite{Bertini2015MFT,BertiniFaggionatoGabrielli2015,KraaijLazarescuMaesPeletier2020,ForastiereRaoEsposito2022,FalascoEsposito2025,BrandnerSaito2020,VanVuSaito2023,Ito2024,ZhongDeWeese2024,WangRen2024,KamijimaItoDechantSagawa2023,SantolinFreitasEspositoFalasco2025}.

\begin{proposition}
\label{Proposition-geometry-topology}
Let \(D\) be the matrix that maps channel currents to the measured transport
records, for example reservoir charge and heat currents. For linear mean
records represented by \(D\), the state dynamics is
thermodynamically complete for these records if and only if
\begin{align}
    D c =0
    \quad
    \text{for every } c\in\ker P .
    \label{Eq-kernel-test}
\end{align}
Equivalently, the rows of \(D\) lie in the row space of \(P\). The number of
independent record directions lost under state-only observation is
\begin{align}
    d_{\mathrm{lost}}
    =
    \operatorname{rank}\left(D|_{\ker P}\right).
    \label{Eq-lost-rank}
\end{align}
\end{proposition}

If \(Dc=0\) for all \(c\in\ker P\), then \(Dj\) is constant on every affine set
of channel currents with the same total ordered-transition currents \(Pj\).
Therefore the record \(Dj\) factors through the state generator. The row-space
condition is the finite-dimensional identity
\((\ker P)^{\perp}=\operatorname{ran}P^{T}\). If there is \(c\in\ker P\) with
\(Dc\neq0\), then \(j\) and \(j+\epsilon c\) give the same state generator and
different transport records. Eq. (\ref{Eq-lost-rank}) is the dimension of the
image of the generator-invisible channel space under the record map \(D\).
This gives the measurable count \(d_{\mathrm{lost}}\) of thermodynamic record
directions that cannot be reconstructed from the state generator.

The criterion can also be used quantitatively. It gives the complete set of
thermodynamic records compatible with a known state generator. Fix the total
ordered-transition currents \(u_{nm}\). For each ordered transition, the
compatible channel currents form a simplex:
\begin{align}
    j_{nm}^{\alpha}
    =
    u_{nm}q_{nm}^{\alpha},
    \qquad
    q_{nm}^{\alpha}\geq0,
    \qquad
    \sum_{\alpha\in{\cal A}_{nm}}q_{nm}^{\alpha}=1 .
    \label{Eq-channel-simplex}
\end{align}
This simplex is the set of all reservoir or filter assignments that give the
same total transition current \(u_{nm}\).

\begin{proposition}
\label{Proposition-compatible-records}
Let \(d_{nm}^{\alpha}\in{\mathbb R}^{q}\) be the vector of recorded charge,
heat, spin, or photon increments carried by channel \(\alpha\) in the
transition \(m\rightarrow n\). For fixed ordered-transition currents \(u\),
define the convex hull of channel increments for that transition by
\begin{align}
    H_{nm}
    =
    \operatorname{conv}
    \left\{
    d_{nm}^{\alpha}:\alpha\in{\cal A}_{nm}
    \right\}.
    \label{Eq-channel-record-hull}
\end{align}
The set \(C(u)\) of all mean record vectors compatible with the same state
generator is
\begin{align}
    C(u)
    =
    \left\{
    \sum_{m\neq n}
    u_{nm}x_{nm}
    :
    x_{nm}\in H_{nm}
    \right\}.
    \label{Eq-compatible-record-set}
\end{align}
For any scalar record \(a\cdot J\), define
\begin{align}
    J_{a}^{-}(u)
    &=
    \sum_{m\neq n}u_{nm}
    \min_{\alpha\in{\cal A}_{nm}}
    a\cdot d_{nm}^{\alpha},
    \nonumber\\
    J_{a}^{+}(u)
    &=
    \sum_{m\neq n}u_{nm}
    \max_{\alpha\in{\cal A}_{nm}}
    a\cdot d_{nm}^{\alpha}.
    \label{Eq-compatible-record-bounds}
\end{align}
The exact state-compatible interval is
\begin{align}
    J_{a}^{-}(u)
    \leq
    a\cdot J
    \leq
    J_{a}^{+}(u).
    \label{Eq-compatible-record-interval}
\end{align}
The record is fixed by the state generator in the direction \(a\) if and only
if \(J_{a}^{-}(u)=J_{a}^{+}(u)\).
\end{proposition}

The proof uses the simplex variables in Eq. (\ref{Eq-channel-simplex}). The
mean record associated with one ordered transition is
\(u_{nm}\sum_{\alpha}q_{nm}^{\alpha}d_{nm}^{\alpha}\), which is exactly
\(u_{nm}H_{nm}\). Summing over ordered
transitions gives Eq. (\ref{Eq-compatible-record-set}). A linear function on
a convex hull reaches its extrema at the listed channel increments, giving
Eqs. (\ref{Eq-compatible-record-bounds}-\ref{Eq-compatible-record-interval}).
The geometry therefore supplies an operational reconstruction bound. From the
state generator and the allowed transport increments one can determine the
full range of heat, particle, or spin records that remain compatible with the
same state dynamics.

The same nullspace controls current noise and full counting statistics, but
the record map must then be understood at the level of the tilted generator.
Let \(\mathcal{L}_{\bm{\chi}}\) denote the off-diagonal channel-resolved tilted
rates in Eq. (\ref{Eq-tilted-offdiag}). For a generator-preserving
perturbation \(c\in\ker P\), the first-order change of the tilted generator is
\begin{align}
    \delta_{c}\mathcal{L}_{\bm{\chi}}
    =
    \sum_{e}
    c_{e}
    e^{\bm{\chi}\cdot d_{e}}
    |n(e)\rangle\langle m(e)|,
    \label{Eq-FCS-null-variation}
\end{align}
where \(e\) runs over directed channels, \(m(e)\) and \(n(e)\) are its initial
and final states, and \(d_{e}\) is the vector of recorded increments. The
channel-resolved tilted generator is fixed by the state generator only if
\(\delta_{c}\mathcal{L}_{\bm{\chi}}=0\) for every \(c\in\ker P\) and for the
counting fields being used. Equivalently, the channel-resolved exponential
weights must be insensitive to all generator-invisible channel
redistributions. If this condition fails and the dominant eigenvalue of the
tilted generator changes, then at least one current cumulant is not determined
by state dynamics. The quantum-dot construction in Sec. \ref{Dot-section}
gives this change explicitly.

The geometric and topological interpretation is therefore operational. The
quotient in Eq. (\ref{Eq-quotient}) gives the fluctuation cost seen after
reservoir-channel assignments have been eliminated, while \(C(u)\) gives the
remaining heat, particle, spin, or photon records compatible with the same
state generator. The discrete connectivity pattern of the transport channels
sets the number of independent redistributions. If there are \(E\)
reservoir-resolved channels and \(E_{0}\) ordered state transitions after
reservoir labels are forgotten, then
\begin{align}
    \dim\ker P
    =
    E-E_{0}.
    \label{Eq-loop-dimension}
\end{align}
The number \(\dim\ker P\) is unchanged by continuous changes of the positive
transition rates. It changes only when a transport channel is added, removed,
or attached to a different ordered transition. Eq. (\ref{Eq-lost-rank}) refines this
connectivity count for thermodynamics: not every generator-invisible
redistribution is visible in a chosen transport measurement, and only the
image of \(\ker P\) under the record map \(D\) produces missing thermodynamic
data. Thus tuning tunnel rates can change the size of the compatible-record
set \(C(u)\), but it cannot remove a nonzero ambiguity unless the channel
connectivity or the measured records are changed. If one knows only an
instantaneous state velocity rather than the full state generator, the larger
nullspace \(\ker(BP)\) also contains closed paths through the reduced state
network. With the full Markovian state generator as input, the relevant hidden
directions are the reservoir-channel redistributions counted by
Eq. (\ref{Eq-loop-dimension}).

For the quantum dot of Sec. \ref{Dot-section}, a rate shift between two
reservoir channels for the same dot transition is an element of \(\ker P\).
If the two reservoirs have different \(\varepsilon_{i}-\mu_{r}\), the heat
record matrix \(D\) does not annihilate that null direction. This gives
precisely the incompleteness shown in Proposition \ref{Proposition-dot}. It
also gives the measurement prescription: state tomography must be supplemented
by enough independent reservoir-current or heat-current measurements to remove
the nonzero directions counted by \(d_{\mathrm{lost}}\).

This prescription can be used without reconstructing every microscopic
channel. Suppose that a set of measured records is represented by the rows of
a matrix \(D_{\mathrm{meas}}\). The remaining ambiguity after state tomography
and these transport measurements is
\begin{align}
    {\cal K}_{\mathrm{rem}}
    =
    \ker P
    \cap
    \ker D_{\mathrm{meas}} .
    \label{Eq-remaining-kernel}
\end{align}
An additional target observable \(D_{\mathrm{tar}}j\) is predictable from the
available data if and only if
\begin{align}
    D_{\mathrm{tar}}c=0
    \quad
    \text{for every }c\in{\cal K}_{\mathrm{rem}} .
    \label{Eq-diagnostic}
\end{align}
Eqs. (\ref{Eq-remaining-kernel}-\ref{Eq-diagnostic}) are the
practical diagnostic. They say which extra transport records must be measured
before a state-only model can be used for thermodynamic prediction. In the
energy-filtered dot, measuring only the total occupation dynamics leaves rate
redistributions among reservoirs in \({\cal K}_{\mathrm{rem}}\). Measuring the
independent reservoir heat currents removes exactly the null directions that
would otherwise change heat noise and heat cross correlations.

\section{Conclusion}
\label{Conclusion-section}

We have shown that a Markovian state master equation can be
thermodynamically incomplete. The failure does not require memory effects,
non-Markovian reservoirs, or an approximate state description. It can occur
inside an ordinary Markovian transport model when the state generator fixes
only total transition rates and not their reservoir-channel assignments.

The main result is a reconstruction criterion. A thermodynamic record is
determined by state dynamics only if it is invariant under every change of
channel assignment that leaves the state generator unchanged. In a
multi-terminal energy-filtered quantum dot, this criterion gives a measurable
prediction: devices with identical occupation dynamics can have different heat
currents, entropy production, and heat-current noise
\cite{SanchezSothmannJordanButtiker2013,WoznyLeijnse2025}. The geometric and
topological analysis explains why. State dynamics is a projection of
channel-current dynamics. The quotient geometry gives the state fluctuations,
while generator-invisible reservoir-channel redistributions carry the missing
thermodynamic records. The compatible-record set in
Eq. (\ref{Eq-compatible-record-set}) turns this statement into a practical
bound on which thermodynamic records remain possible after the state generator
has been measured.

The practical limitation is direct: a master equation fitted from state
tomography may be sufficient for relaxation and stationary occupations, but
insufficient for thermodynamic performance and noise. To make thermodynamic
predictions complete, one must specify not only the state generator but also
the reservoir and measurement records associated with the transitions
\cite{Seifert2019Inference,ErtelSeifert2024,PtaszynskiEsposito2023}.

\bibliographystyle{apsrev4-2}
\bibliography{Main}

\end{document}